\def\papertitle{Differentiable Attenuation Filters for Feedback Delay Networks}
\def\paperauthorA{Ilias Ibnyahya}
\def\paperauthorB{Joshua D. Reiss}
\documentclass[twoside,a4paper]{article}
\usepackage{etoolbox}
\usepackage{dafx_25}
\usepackage{amsmath,amssymb,amsfonts,amsthm,booktabs, bm}
\usepackage{euscript}
\usepackage[T1]{fontenc}
\usepackage[utf8]{inputenc}
\usepackage{ifpdf}
\usepackage[english]{babel}
\usepackage{caption}
\usepackage{subfig} % or can use subcaption package
\usepackage{color}

\input glyphtounicode
\ninept

\newcounter{numauth}
\newcounter{listcnt}
\newcommand\authcnt[1]{\ifdefined#1 \stepcounter{numauth} \fi}

\newcommand\addauth[1]{
\ifdefined#1 
\stepcounter{listcnt}
\ifnum \value{listcnt}<\value{numauth}
\appto\authorslist{, #1}
\else
\appto\authorslist{~and~#1}
\fi
\fi}
\authcnt{\paperauthorB}
\authcnt{\paperauthorC}
\authcnt{\paperauthorD}
\authcnt{\paperauthorE}
\authcnt{\paperauthorF}
\authcnt{\paperauthorG}
\authcnt{\paperauthorH}
\authcnt{\paperauthorI}
\authcnt{\paperauthorJ}
\def\authorslist{\paperauthorA}
\addauth{\paperauthorB}
\addauth{\paperauthorC}
\addauth{\paperauthorD}
\addauth{\paperauthorE}
\addauth{\paperauthorF}
\addauth{\paperauthorG}
\addauth{\paperauthorH}
\addauth{\paperauthorI}
\addauth{\paperauthorJ}
\usepackage{times}
\newif\ifpdf
\ifx\pdfoutput\relax
\else
   \ifcase\pdfoutput
      \pdffalse
   \else
      \pdftrue
   \fi
\fi

\ifpdf % compiling with pdflatex
  \usepackage[pdftex,
    pdftitle={\papertitle},
    pdfauthor={\authorslist},
    pdfsubject={Proceedings of the 28th International Conference on Digital Audio Effects (DAFx25)},
    colorlinks=false, % links are activated as color boxes instead of color text
    bookmarksnumbered, % use section numbers with bookmarks
    pdfstartview=XYZ % start with zoom=100% instead of full screen; especially useful if working with a big screen :-)
  ]{hyperref}
  \usepackage[pdftex]{graphicx}
\else % compiling with latex
  \usepackage[dvips]{epsfig,graphicx}
  \usepackage[dvips,
    pdftitle={\papertitle},
    pdfauthor={\authorslist},
    pdfsubject={Proceedings of the 28th International Conference on Digital Audio Effects (DAFx25)},
    colorlinks=false, % no color links
    bookmarksnumbered, % use section numbers with bookmarks
    pdfstartview=XYZ % start with zoom=100% instead of full screen
  ]{hyperref}
\fi
\usepackage[hypcap=true]{caption}
\title{\papertitle}

\affiliation{
\paperauthorA\, and \paperauthorB ,\thanks{\vspace{-3mm}}}
{\href{https://www.c4dm.eecs.qmul.ac.uk/}{Centre for Digital Music} \\ Queen Mary University of London \\ London, UK\\
{\tt \href{mailto:i.ibnyahya@qmul.ac.uk}{i.ibnyahya@qmul.ac.uk} | \href{mailto:joshua.reiss@qmul.ac.uk}{joshua.reiss@qmul.ac.uk}}
}

\begin{document}
% more pdf-tex settings:
\ifpdf % used graphic file format for pdflatex
  \DeclareGraphicsExtensions{.png,.jpg,.pdf}
\else  % used graphic file format for latex
  \DeclareGraphicsExtensions{.eps}
\fi

%\makeatletter
%\pdfbookmark[0]{\@pdftitle}{title}
%\makeatother

\maketitle

% \begin{abstract}
% We propose a novel approach for fitting Second Order Sections (SOS) of Infinite Impulse Response (IIR) digital filters to attenuation filters commonly used in digital audio reverberation with Feedback Delay Networks (FDN). Attenuation filters typically control the frequency-dependent reverberation time in reverberation synthesis. In this work, we leverage principles of biquadratic filter design and gradient descent optimization to achieve an accurate and scalable filter response while maintaining a regular parametric equalizer parameter set. A key advantage of our approach is its scalability to a large number of filters, as the frequency , gain and Quality Factor (Q) parameters remain common across parallel delay lines. This reduces the number of optimization parameters by a factor of two compared to existing methods. The result is a simpler, differentiable framework that can be efficiently applied to attenuation filters in digital reverberation.
% \end{abstract}

\begin{abstract}
We introduce a novel method for designing attenuation filters in digital audio reverberation systems based on Feedback Delay Networks (FDNs). Our approach uses Second Order Sections (SOS) of Infinite Impulse Response (IIR) filters arranged as parametric equalizers (PEQ), enabling fine control over frequency-dependent reverberation decay. Unlike traditional graphic equalizer designs, which require numerous filters per delay line, we propose a scalable solution where the number of filters can be adjusted. The frequency, gain, and quality factor (Q) parameters are shared parameters across delay lines and only the gain is adjusted based on delay length. This design not only reduces the number of optimization parameters, but also remains fully differentiable and compatible with gradient-based learning frameworks. Leveraging principles of analog filter design, our method allows for efficient and accurate filter fitting using supervised learning. Our method delivers a flexible and differentiable design, achieving state-of-the-art performance while significantly reducing computational cost.
\end{abstract}

\section{Introduction}
\label{sec:intro}

Feedback Delay Networks (FDNs) are a common technique in artificial reverberation, used to simulate the reflections found in real acoustic environments \cite{valimaki_fifty_2012, jot_analysissynthesis_1992}. They are especially used for modeling the late reflections of a reverberant field, where their scalable structure allows an effective balance between computational efficiency and acoustic accuracy. In typical FDNs, attenuation filters are employed to control the frequency-dependent decay, while the feedback matrix is designed to be both lossless and colorless \cite{prawda_improved_2019, santo_differentiable_2023}. Each feedback path incorporates an absorption filter that gradually reduces energy over time. By using frequency-dependent filters, the reverberation decay can be precisely shaped across the spectrum, offering perceptually meaningful control over the reverberation time ($T_{60}$).

Attenuation filters, also referred to as absorption filters, simulate the loss of energy at different frequencies. While early methods relied on simple low-pass filters \cite{moorer_about_1979}, more advanced designs now use graphic equalizers (GEQs) for finer spectral resolution \cite{jot_proportional_2015, valimaki_accurate_2017, valimaki_two-stage_2024}. Some recent work also explores deep learning approaches to directly learn IIR filter coefficients \cite{lee_differentiable_2022}, but these methods typically lack a clear mechanism for uniformly controlling the decay time across all frequency bands. As a result, they may produce filters with inconsistent spectral decay characteristics, making them less suitable for applications where precise and homogeneous reverberation control is required.
Traditional Graphic Equalizers (GEQs) present limitations when integrated into Feedback Delay Networks (FDNs). Their typically limited attenuation range can result in poor matching within long delay lines. Furthermore, the inherent characteristics of the bell filters used in GEQs can introduce frequency response mismatches at both DC and Nyquist frequencies, as these filters must reach unity gain at the spectrum's edges \cite{liski_quest_2017}. The two-stage attenuation filter (TSAF) introduced in \cite{valimaki_two-stage_2024} combines a shelving filter with a 31-band GEQ to mitigate these issues and achieves strong performance with a low relative $T_{60}$ error on real-world impulse responses \cite{prawda_calibrating_2022}. Although a reduced version with 12 bands lowers computational cost, it can compromise accuracy. Furthermore, the granularity of these methods is tied to the number of bands per octave, making parametric equalizers a more computationally efficient alternative.
Despite significant progress, achieving a fully differentiable and trainable FDN remains a challenge. Prior research has succeeded in optimizing gain values and feedback matrices \cite{santo_differentiable_2023, santo_rir2fdn_2024}, but attenuation filters remain difficult to optimize in a differentiable manner \cite{valimaki_two-stage_2024, lee_differentiable_2022}. The challenge lies in the high dimensionality of the parameter space, as each delay line requires multiple filters and there is lack of a direct mapping between reverberation time and filter coefficients. Existing approaches often use fixed frequency and Q values, optimizing only the gain, and require a complex multi-stage processes. As such, they are not ideal for training in end-to-end differentiable frameworks. Another practical issue concerns the accuracy of digital filters across the full audio spectrum. Filter designs based on the bilinear transform, such as those in \cite{nercessian_neural_2020}, tend to lose accuracy near the Nyquist frequency. Improved formulations like those of \cite{liski_graphic_2019, john_improving_2018} and the symmetric shelving design of \cite{valimaki_all_2016} help maintain better precision, particularly in the high-frequency range.

To address these challenges, we propose a method that uses a variable amount of biquad filters per delay line, structured as a parametric equalizer with two shelving filters and bell filters. This setup delivers a magnitude response comparable to the more complex TSAF with 12 bands while reducing the number of filters by more than two-thirds. For an 8th-order FDN with a 12-bands PEQ, our method requires 96 biquads compared to 248 for a 31-band GEQ. Our system is fully differentiable and supports optimization through gradient descent, allowing all filter parameters to be trained jointly. This enables efficient and scalable filter fitting without compromising performance, which is particularly advantageous in resource-constrained environments such as embedded systems. An implementation of our approach is available on GitHub \footnote{\url{https://github.com/ilias-audio/iir_match}}.

\section{METHOD}
\subsection{Parametric Equalizer Design}
\begin{figure}[ht]
    \centering
    \includegraphics[width=\columnwidth]{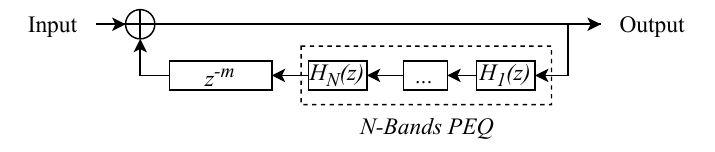}
    \caption{\label{delay_filter}{\it Recursive filter design with an N-band parametric equalizer.}}
\end{figure}

The parametric equalizer (PEQ) used in this work is scalable. It consists of three types of second-order filters: bell, low-shelf, and high-shelf. A typical configuration for digital reverberation is illustrated in Figure~\ref{delay_filter}. The number of bands \(N\) and the delay line lengths \(m\) are variable. The first and last filters, \(H_1(z)\) and \(H_N(z)\), are implemented as low-shelf and high-shelf filters respectively, while the intermediate filters \(H_2(z)\) through \(H_{N-1}(z)\) are bell filters.

Each filter is defined by three parameters: gain \(G\), center frequency \(f_c\), and quality factor \(Q\). The gain is expressed in dB and converted to a linear amplitude factor \(A\) as follows:

\begin{equation}
    A = \sqrt{10^{\frac{G}{20}}} = 10^{\frac{G}{40}}
\end{equation}

We base our analog filter design on the general second-order prototype:

\begin{equation}
    H(s) = g_{\infty} \frac{s^2 + \frac{\omega_z}{Q_z}s + \omega_z^2}{s^2 + \frac{\omega_p}{Q_p}s + \omega_p^2}
\label{eq:LaplaceBiquad}
\end{equation}

In this expression, \(g_{\infty}\) is the high-frequency gain, while \(\omega_z, Q_z\) and \(\omega_p, Q_p\) define the zeros and poles respectively. Laplace domain transfer functions from \cite{bristow-johnson_audio-eq-cookbook_2021} are used to define the response of Bell $(H_B)$, Low-Shelf $(H_{LS})$ and High-Shelf $(H_{HS})$ filters:

\begin{equation}
    H_B(s) = \frac{s^2 + \frac{A}{Q}s + 1}{s^2 + \frac{s}{AQ} + 1}
\label{eq:LaplaceBell}
\end{equation}

\begin{equation}
    H_{LS}(s) =  A \frac{s^2 + \frac{\sqrt{A}}{Q}s + A}{As^2 + \frac{\sqrt{A}}{Q}s+ 1}
\label{eq:LaplaceLowShelf}
\end{equation}

\begin{equation}
    H_{HS}(s) = A \frac{As^2 + \frac{\sqrt{A}}{Q}s+ 1}{s^2 + \frac{\sqrt{A}}{Q}s + A}
\label{eq:LaplaceHighShelf}
\end{equation}

To evaluate the frequency response from the Laplace domain we substitute $s=j \frac{2\pi f}{2\pi f_c}$ and express the frequency response as a function of frequency $f$, with the center frequency $f_c$, linear gain $A$, and Q-factor $Q$ as parameters:

\begin{equation}
    \left|H_B(f)\right| = \sqrt{\frac{(1 - \frac{f}{f_c}^2)^2 + (\frac{A\frac{f}{f_c}}{Q})^2}{(1 - \frac{f}{f_c}^2)^2 + (\frac{\frac{f}{f_c}}{AQ})^2}}
\label{eq:MagBell}
\end{equation}

\begin{equation}
   \left| H_{LS}(f)\right| = A \sqrt{\frac{(A - \frac{f}{f_c}^2)^2 + (\frac{\sqrt{A}\frac{f}{f_c}}{Q})^2}{(1 - A\frac{f}{f_c}^2)^2 + (\frac{\sqrt{A}\frac{f}{f_c}}{Q})^2}}
\label{eq:MagLowShelf}
\end{equation}

\begin{equation}
    \left|H_{HS}(f)\right| = A \sqrt{\frac{(1 - A\frac{f}{f_c}^2)^2 + (\frac{\sqrt{A}\frac{f}{f_c}}{Q})^2}{(A - \frac{f}{f_c}^2)^2 + (\frac{\sqrt{A}\frac{f}{f_c}}{Q})^2}}
\label{eq:MagHighShelf}
\end{equation}

The magnitude response of the full PEQ is obtained by summing the logarithmic magnitude of each band:

\begin{equation}
20\log_{10}\left| H_{PEQ}(f) \right| = \sum_{i=1}^{N} 20\log_{10}\left| H_i(f) \right|
\label{eq:EQMAg}
\end{equation}

Each filter band thus requires optimization over three parameters. The analog prototype is converted to a digital filter using the method in \cite{liski_graphic_2019}.

Following the approach of Prawda et al.~\cite{prawda_improved_2019}, we model the attenuation response to scale with the delay line length in samples. The target magnitude for a given delay line \(k\) of length \(m_k\) and sample rate $f_s$ becomes:

\begin{equation}
20\log_{10}|H_{PEQ_k}(f)| = \frac{-60\ m_k}{T_{60}(f)\ f_s}
\label{eq:T60Mag}
\end{equation}

This relation ensures that all filters can share a common set of parameters while scaling the gain according to delay line length. We define the slope as:

\begin{equation}
\gamma(f) = \frac{-60}{T_{60}(f)}
\end{equation}

and the corresponding gain per delay line is:

\begin{equation}
G(f) = \frac{\gamma(f)\ m_k}{f_s}
\end{equation}

Using this formulation, the PEQ can be shared across all delay lines with only the gain scaling changing, significantly reducing the number of parameters.

\subsection{Parameter Optimization}

To optimize the filter parameters, we frame the task as supervised regression. The objective is to match the frequency-dependent target magnitude derived from a desired \(T_{60}(f)\). The loss function is the mean squared error (MSE) between the target and the PEQ response:

\begin{equation}
    \text{MSE} = \frac{1}{N} \sum_{i=1}^{N} \left( \frac{-60\ m_i}{T_{60}(f)\ f_s} - 20\log_{10}|H_{PEQ}(f)| \right)^2
\end{equation}

We implement the full model in PyTorch, where each filter band is differentiable and parameterized by frequency, gain, and Q-factor. Optimization is performed with the Adam optimizer \cite{kingma_adam_2015}, using an initial learning rate of 0.1 and 10,000 iterations. These values are empirical but provide enough iterations for complex cases with more than 12 bands.

At each step, gradients are computed with respect to all parameters via backpropagation, allowing efficient training of the full PEQ. The use of shared parameters across delay lines ensures scalability and enables the application of our method even in embedded systems.

Finally, we validate the method using a dataset of 1000 RIRs \cite{prawda_calibrating_2022}, following the evaluation protocol from \cite{valimaki_two-stage_2024}, and compare it against state-of-the-art approaches.

\section{Results}

\subsection{Reverberation Time Evaluation}

We evaluated the precision of our parametric equalizer designs against a dataset of measured room impulse responses (RIR) in a variable acoustics environment \cite{prawda_dataset_2022}. The reverberation times are estimated in third-octave frequency bands over 31 bands. A subset of 1000 RIRs is used, following a comparable method to the one exposed in \cite{valimaki_two-stage_2024}. We linearly interpolate the 31 bands of RIR measurements to 512 logarithmically spaced points between DC and Nyquist. Our approach is tested by randomly generating delay line lengths between 0.01 and 0.3 seconds to emulate the behavior of different types of rooms and halls. This strategy also allows for evaluation against a wide range of attenuation filter combinations, where, for example, an increase in delay corresponds to greater attenuation. Each test case involved matching the frequency dependent reverberation time (\(T_{60}(f)\)) with the optimized magnitude response of the filter converted back to a reverberation time (\(\hat{T}_{60}(f)\)). 

\begin{equation*} T_\text {60}{_{\mathrm{,\,error}}} = \frac{T_\text {60}(f) - \hat{T}_\text {60}(f)}{T_\text {60}(f)} \times 100\%. \tag{9} \end{equation*}

\begin{figure}[t]
\centerline{\includegraphics[scale=0.8]{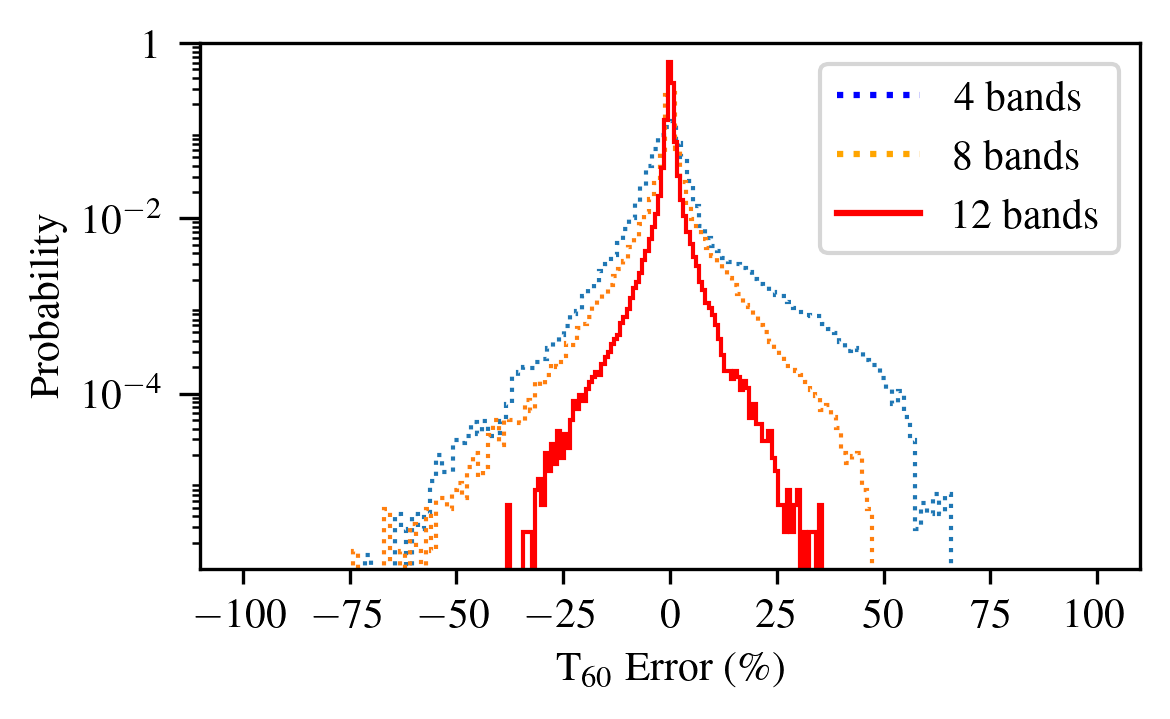}}
\caption{\label{distribution}{\it Reverberation time relative error of three different PEQ designs applied to a 1000 RIR reverberation time measurements dataset with a distribution of different delay lengths.}}
\end{figure}

We calculate the error $T_{60,{error}}$ for each point of the interpolated dataset and for each of the 1000 reverberation time estimations. This effectively means we evaluate the error against 512,000 data points. We then use this relative error to determine the error distribution (Figure~\ref{distribution}), observing that high levels of inaccuracy are avoided. For instance, the smallest equalizer (EQ) yields an inaccuracy within $\pm 25\%$.12 bands PEQ shows a narrow distribution of the error centered around zero. Adding more bands makes the resulting frequency reverberation time attenuation more accurate.

In Figure~\ref{median}, even with as few as 4 bands, the model captures the general trend of the target magnitude curve. Increasing the number of bands to 8 and 12 further reduces the spread of the error, bringing the performance close to that of the state-of-the-art Two-Stage Attenuation Filter (TSAF) with 31 bands. This highlights the effectiveness of our differentiable approach in capturing complex reverberation behavior with far fewer filters.

\subsection{Attenuation Filter Magnitude Accuracy}

\begin{figure}[t]
\centerline{\includegraphics[scale=0.8]{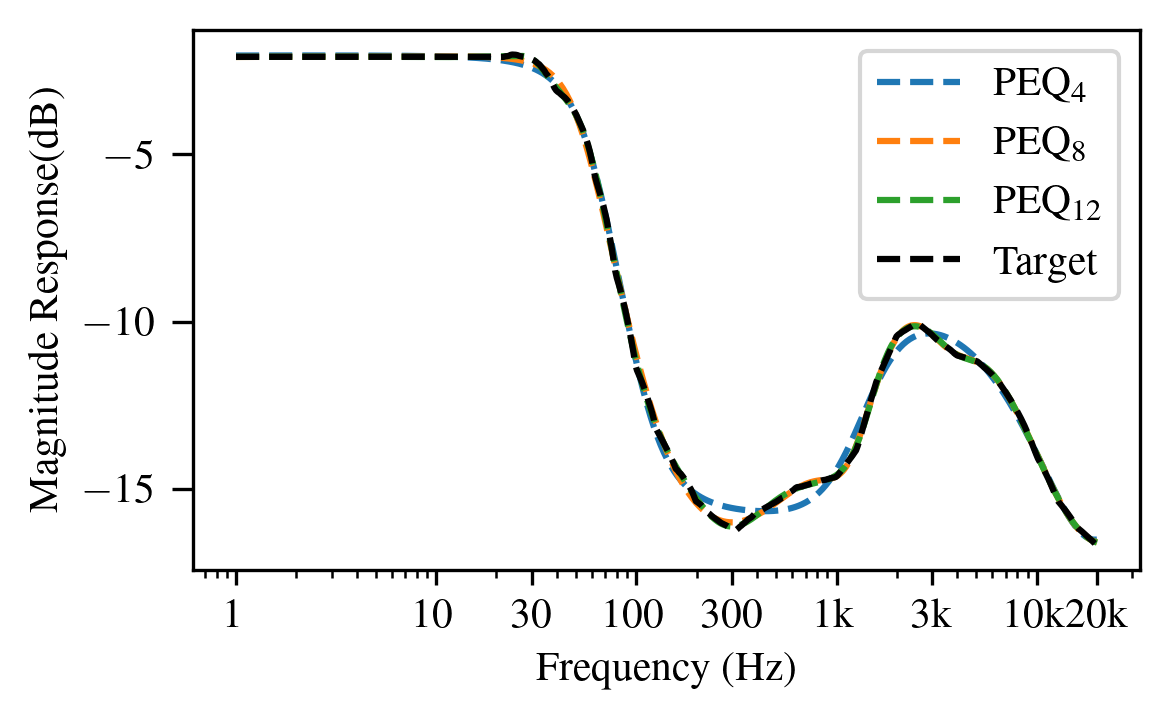}}
\caption{\label{median}{\it Magnitude response accuracy for different PEQ designs based on the dataset median reverberation time at each frequency.}}
\end{figure}

To assess the fidelity of the frequency response, we measured the deviation between the optimized equalizer response and the median target $T_{60}$ curve across the dataset. We used a delay line of 100 milliseconds and a sampling rate of 48kHz. Figure~\ref{median} displays the achieved magnitude response for various PEQ configurations.

The results show that even low-order designs (e.g., 4 or 8 bands) provide a smooth and perceptually relevant approximation of the target response. The 12-band PEQ closely tracks the target curve, exhibiting only minor deviations, and is nearly indistinguishable from the target.

\begin{table}[ht]
    \centering
    \begin{tabular}{ccc|cc}
        \toprule
        Type & $MSE(x,\hat{x})$ & $MAE(x,\hat{x})$ & $OP$ & $P$ \\
        \midrule
        $PEQ_4$      & $4.8 \times 10^{-2}$   & $6.3 \times 10^{-1}$ & $36$  & $12$ \\
        $PEQ_8$      & $7.3 \times 10^{-3}$   & $4.1 \times 10^{-1}$ & $72$  & $24$ \\
        \midrule
        $PEQ_{12}$   & \bm{$1.7 \times 10^{-3}$}   & $2.2 \times 10^{-1}$ & \bm{$108$} & $36$ \\
        $TSAF_{31}$  & $1.9 \times 10^{-3}$   & \bm{$1.2 \times 10^{-1}$} & $284$ & \bm{$33$} \\
        \bottomrule
    \end{tabular}
    \caption{\textit{Magnitude Mean-Squared Error (MSE) and Maximum Absolute Error (MAE) in dB between the target response ($x$) and the predicted response ($\hat{x}$) using N-band PEQs and the 31-band TSAF. OP denotes the number of arithmetic operations, and P the number of trainable parameters.}}

    \label{tb:Median Error}
\end{table}
Quantitative metrics are presented in Table~\ref{tb:Median Error}, comparing the Mean Squared Error (MSE), Maximum Absolute Error (MAE), and computational cost (additions and multiplications, noted as OP) for each method using the same delay length and sampling rate. Additionally, P represents the number of trainable parameters, providing insight into the complexity and memory footprint of each configuration. Our proposed 12-band PEQ achieves a similar MSE to the TSAF while using only $38\%$ of the operations and maintaining a comparable number of parameters (36 for our approach and 33 for $TSAF_{31}$), indicating a favorable balance between performance, computational efficiency, and model size for the proposed architecture. These findings confirm that our differentiable PEQ design offers a balance between performance and complexity, making it suitable for real-time applications.

\section{Conclusions}

We presented a simple and efficient method for designing attenuation filters in Feedback Delay Networks using a differentiable parametric equalizer framework. Our approach uses fewer filters than traditional methods while maintaining comparable accuracy and stability, making it well-suited for real-time and embedded applications. By optimizing the filters with gradient descent and working on continuous target curves, we avoid the need for complex neural networks. 
\\
\\
Future work should explore integrating this method into a complete FDN architecture and evaluate its effectiveness using biquad coefficients. This would help assess its performance in realistic audio processing pipelines and confirm its suitability for end-to-end optimization tasks.

%\newpage

\bibliographystyle{IEEEbib}
\bibliography{Zotero}
\end{document}